\documentclass[preprintnumbers,amsmath,amssymb,twocolumn]{revtex4}
\usepackage{graphicx}% Include figure files
\usepackage{dcolumn}% Align table columns on decimal point
\usepackage{bm}% bold math
\usepackage{natbib}
\usepackage[caption=false]{subfig}

\begin{document}

\title{Strong Coupling between a Single NV Spin and the Rotational Mode \\ of Diamonds Levitating in an Ion Trap}

\author{T. Delord$^{1}$}
\author{L. Nicolas$^{1}$}
%%\author{L. Schwab$^{1}$}
\author{Y. Chassagneux$^{1}$} 
\author{G. H\'etet$^{1}$} 

\affiliation{$^1$Laboratoire Pierre Aigrain, Ecole normale sup\'erieure, PSL Research University, CNRS, Universit\'e Pierre et Marie Curie, Sorbonne Universit\'es, Universit\'e Paris Diderot, Sorbonne Paris-Cit\'e, 24 rue Lhomond, 75231 Paris Cedex 05, France.}

\begin{abstract}
A scheme for strong coupling between a single atomic spin and the rotational mode of levitating nanoparticles is proposed. The idea is based on spin read-out of NV centers embedded in aspherical nanodiamonds levitating in an ion trap. We show that the asymmetry of the diamond induces a rotational confinement in the ion trap. Using a weak homogeneous magnetic field and a strong microwave driving we then demonstrate that the spin of the NV center can be strongly coupled to the rotational motion of the diamond.
\end{abstract}
\maketitle

Experiments in the field of opto-mechanics showed control of macroscopic mechanical oscillators very close to their ground state of motion \cite{Aspelmeyer}. These accomplishments provide great opportunities to observe quantum superpositions with macroscopic systems. Although progress are being made with room temperature oscillators \citep{reinhardt2016ultralow,norte2016mechanical,tsaturyan2016ultra}, the difficulty in most experiments is that they require cooling of the oscillators down to milliKelvin temperatures or carefully ingineered nano-mechanical oscillators 
%It is indeed %technically challenging to decouple them from the thermal reservoir in order to be well within the quantum regime at room temperature, simply 
 because they are clamped to a structure. Inspired by ideas for mechanical control of oscillating cantilevers using magnetic field sensitive probes \cite{Treutlein, Rabl, Arcizet, Kolkowitz,li2016hybrid}, trapped macroscopic objects coupled to single spins via magnetic field gradients  are envisioned \cite{yin}. There, the mechanical support is completely removed so one could operate at room temperature and reach high quality factors \cite{Chang19012010}. 
Further, the spins coupled to the massive object can be used to create matter wave interference \cite{Scala} and Schr${\rm \ddot o}$dinger cat states where the spin is entangled with the collective oscillator motion \cite{yin,yin2013optomechanics}. 

Many experimental protocols are being explored to couple the center of mass mode of levitating objects to single spins, most of which use diamonds with embedded Nitrogen Vacancy (NV) centers in dipole traps \cite{yin}. In recent experiments however, despite the mechanical support being completely removed, light scattering from the optically levitated object significantly alters the photophysical properties of the NV centers \cite{Neukirch, Horowitz, Geiselmann}. Although advances have been made in this direction \cite{Frangeskou}, many groups indeed observe strong heating at low vacuum pressures which quenches the NV photoluminescence \cite{Rahman,Neukirch, Hoang}. On the other hand, scattering free traps such as Paul traps or magneto-gravitational traps allow reaching lower vacuum \cite{vacuumESR,Hsu} although currently with a lower trapping frequency. One further difficulty with the hybrid proposals is that reaching strong coupling between a single spin and the center of mass mode implies high magnetic field gradients in the range of $10^5$ to $10^7$~T/m \cite{Rabl, yin2013optomechanics}, which is very challenging.

In this paper, we present a scheme for strong coupling between a single spin and levitating nanoparticles that leverages most of these issues. First, we propose using a Paul trap for rotational confinement of charged aspherical nanodiamonds.  Second, the rotational degree of freedom is coupled to the spin of embedded NV centers via homogeneous magnetic fields. 
The proposal makes use of the inherent quantization axis of the NV center together with the sensibility of its spin energy levels to the magnetic field. We show that homogeneous magnetic fields in the range of tens of milliteslas are enough to enter the strong coupling regime for the rotational mode of prolate particles.

\begin{figure}[ht!]
\centerline{\scalebox{0.2}{\includegraphics{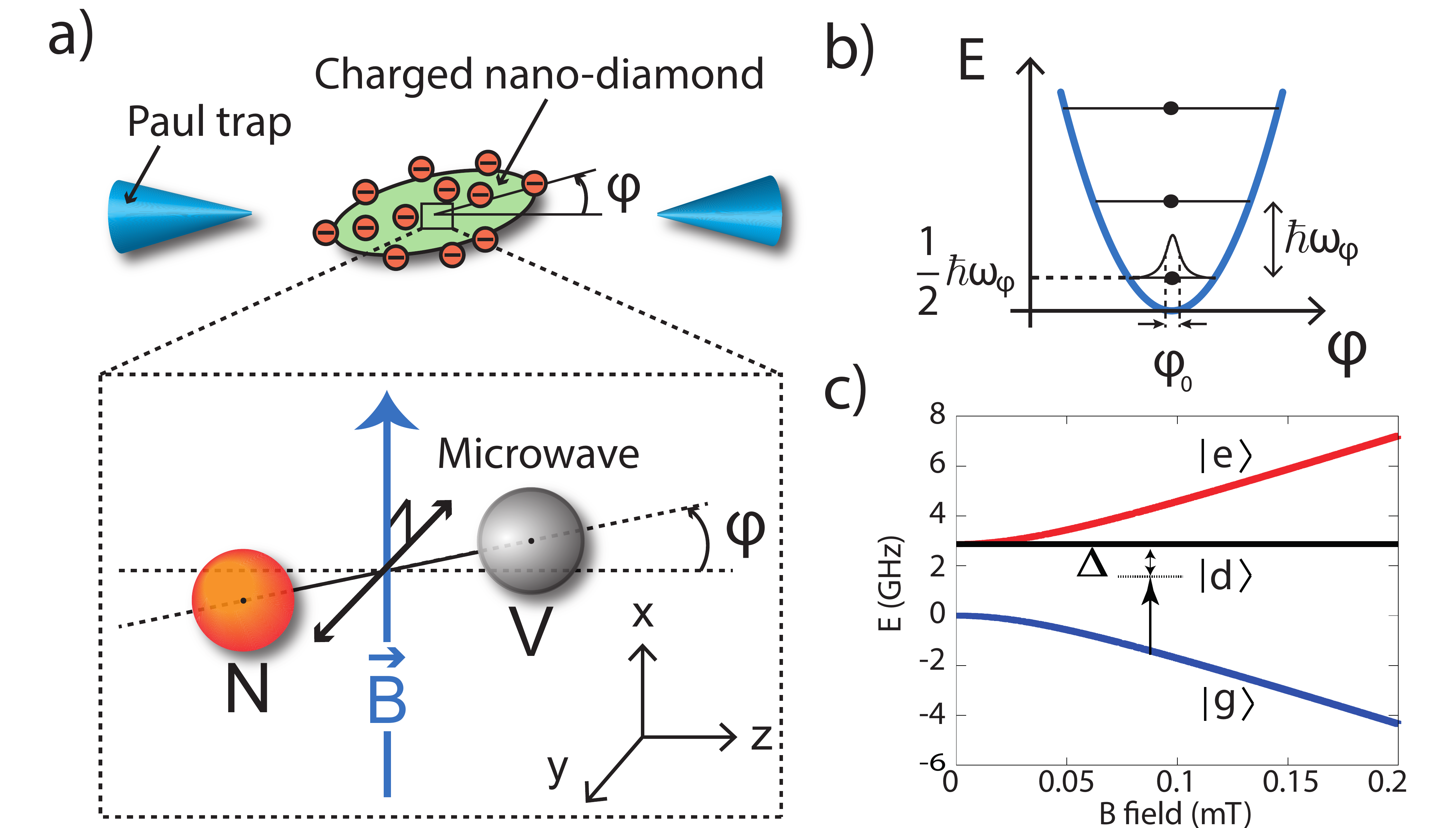}}}
\caption{a) Schematics showing a prolate nanodiamond levitating in a needle Paul trap. The spin of a nitrogen vacancy center inside the diamond senses the rotation of the particle in the presence of a transverse B-field. b) Harmonic potential energy as a function of the angle $\phi$ between the particle and the main trap axis, with rotational frequency $\omega_\phi$. c) NV center ground state level shifts in the presence of a transverse magnetic field. The arrow shows the microwave driving with a detuning $\Delta$ from the $|g\rangle$ to $|d\rangle$ transition. }\label{setup}
\end{figure}

Fig. 1 shows a schematics of the proposal. A prolate diamond is levitating in a needle Paul trap.  
The coupling between the diamond rotational mode and the NV center relies on the control of its electronic spin in a homogeneous magnetic field. The NV centers in diamond consist of a substitutional nitrogen atom (N) associated with a vacancy (V) in an adjacent lattice site of the diamond matrix. This defect behaves as an artificial atom trapped in the diamond matrix and exhibits a strong photoluminescence in the red which allows the detection of individual NV defects at room temperature.
It is also possible to optically initialize and read-out the electronic spin of the NV center thanks to the presence of a metastable level and an intersystem crossing \cite{Gruber}. 
Compared to single atoms where the quantization axis is defined with respect to the B-field, with NV centers, the ground state spin-spin interaction sets a preferential quantization axis, namely the N-V direction, as shown in the inset. This feature is the cornerstone of this proposal.

\section{Rotational confinement in a Paul trap \label{angularconfinement}}
NV centers were detected with diamonds levitating in a Paul trap in \cite{Kuhlicke, delord2016}. 
In \citep{delord2016}, the electronic spin resonance of nanodiamonds was further employed to experimentally demonstrate their angular stability.  Here, we show that the rotation about two axes is ruled by a Mathieu equation so that the angle is stabilized, like the center of mass.  To show this, let us consider the following time-dependent quadratic electric potential:
$V_E(t)=\frac{\eta V(t)}{z_0^2} \left( z ^2 - \frac{1}{2}x^2-\frac{1}{2}y^2 \right),$
where  $V(t)=V_{\rm dc}+V_{\rm ac} \cos(\Omega t)$ is the voltage applied to the needle electrodes oscillating at a frequency $\Omega/2\pi$, $z_0$ is the distance between the two needles and $\eta$ an efficiency parameter that accounts for deviations from ideal hyperbolic electrode shapes \cite{Pau90}. To evaluate the rotational frequency, we take a particle with total surface charge $Q$, and assume that the charge centroid coincides with the center of mass at all times. 

One can then calculate the torque applied by the electric field to the particle. For an element of surface $dS$ with charge $dQ$ and in the fixed $xyz$ frame the torque reads 
\begin{eqnarray}\vec{dM}=\frac{\eta V(t)}{z_0^2}\begin{pmatrix}
-3yx\\
3xz\\
0\\
\end{pmatrix} dQ.
\end{eqnarray}

This torque can then integrated over the whole surface of the particle to obtain Euler's rotation equations. To factorize the dependency on the orientation of the particle, the integration is done in the rotating frame $XYZ$ whose axes are fixed to the particle and parallel to its principal axes of inertia. We consider the particle to be symmetric about its $Z$ axis and hence use only two Euler angle $\phi_1$, $\phi_2$ to define the $XYZ$ frame : $\phi_1$ for a first rotation of the initial frame $xyz$ about the $y$ axis and $\phi_2$ for a second rotation of the rotated frame $x'y'z'$ about the rotated $x'$ axis. The matrix allowing one to obtain $XYZ$ from $xyz$ is : \begin{eqnarray}
R(\phi_1,\phi_2)= \begin{pmatrix}
\cos\phi_1 & 0 & \sin \phi_1 \\
0 & 1 & 0 \\
-\sin\phi_1 & 0 & \cos\phi_1 \\
\end{pmatrix}\begin{pmatrix}
1 & 0 & 0 \\
0 & \cos\phi_2 & -\sin\phi_2 \\
0 & \sin\phi_2 & \cos\phi_2
\end{pmatrix}
\end{eqnarray}

After changing the basis to integrate over the surface of the particle we find the total torque along the $X$ and $Y$ axis to be : 

\begin{equation}\begin{array}{l l c}
M_X&=&\frac{3 \eta V}{2 z_0^2}  \cos^2\phi_1\sin(2\phi_2) \iint (Z^2-Y^2)dQ \\
&&\rm and\\
M_Y&=&\frac{3\eta V}{2 z_0^2}\cos\phi_2 \sin(2\phi_1) \iint (Z^2-X^2)dQ. \\
\end{array}
\end{equation}
In the limit of small angles $\phi_1 , \phi_2 \ll 1$, the Euler equations for the angles $\phi_{1,2}$ are therefore become 
\begin{equation}\begin{array}{l l}
I_x \ddot{\phi_1}&-V(t)\big[  \frac{3\eta}{z_0^2} \iint (Z^2-X^2)dQ \big] \phi_1=0\\
I_y \ddot{\phi_2}&-V(t)\big[ \frac{3\eta}{z_0^2} \iint (Z^2-Y^2)dQ \big] \phi_2=0,\end{array}
\end{equation}
where $I_{x,y}$ are the moment of inertia relative to $x$, resp. $y$ (see part I of SI).
Eqs. (1) are Mathieu equations for the angles $\phi_1$, $\phi_2$. Within their stability conditions, they yield a harmonic confinement for both rotation angles of the particle, at secular frequencies :
\begin{equation}\label{eqconfinement}
\omega_\mu = \frac{\Omega}{2} \sqrt{a_\mu+\frac{q_\mu^2}{2}},
\end{equation}
with dimensionless parameters :
\begin{equation}\begin{array}{l c r}
q_\mu&=& 3 \frac{Q S_\mu}{I_\mu}\frac{V_{ac}}{z_0^2}\frac{1}{\Omega^2}~~{\rm and}~~
a_\mu=-6 \frac{Q S_\mu}{I_\mu}\frac{V_{dc}}{z_0^2}\frac{1}{\Omega^2}
,\end{array}
\end{equation} 
where $\mu=X,Y$, $S_X=R_Z^2-R_Y^2$, $S_Y=R_Z^2-R_X^2$ and $R_\mu^2=\iint \mu^2dQ/Q$. At this stage, the calculations do not assume a homogeneous charge distribution. 

Note that a small angle can initially be reached through damping of the surrounding gas as observed in \citep{delord2016} at atmospheric pressure or through parametric feedback under vacuum. Once within the small angle approximation, the angular extension of the harmonic oscillator is due to the temperature of the surrounding gas and can be calculated using the equipartition of the energy. For the prolate particles of 20 nm and 80 nm diameter as described bellow, we find $\sqrt{\langle \phi^2 \rangle}=0.16$ and $0.05$ rad.

The rotational confinement not only depends on the charge to mass ratio and on the generated potential, like for the center of mass, but also crucially on the geometry of the particle. The factor $Q S_\mu/I_\mu$ in Eq. (3), which can be written $\iint (Z^2-X^2)dQ/ \left(\iiint Y^2 dm\right)$ for $\mu=Y$, can indeed be increased substantially using an asymmetric particle. A more significant advantage of using the rotational mode however comes from the possibility to strongly couple to a single spin via a weak homogeneous magnetic field.
\captionsetup[subfigure]{position=top,labelformat=empty,singlelinecheck=off,justification=raggedright}

\begin{figure}[ht!]
\centering
\subfloat[\label{coolingmapB}\label{spec}]{\includegraphics[width=0.5\textwidth]{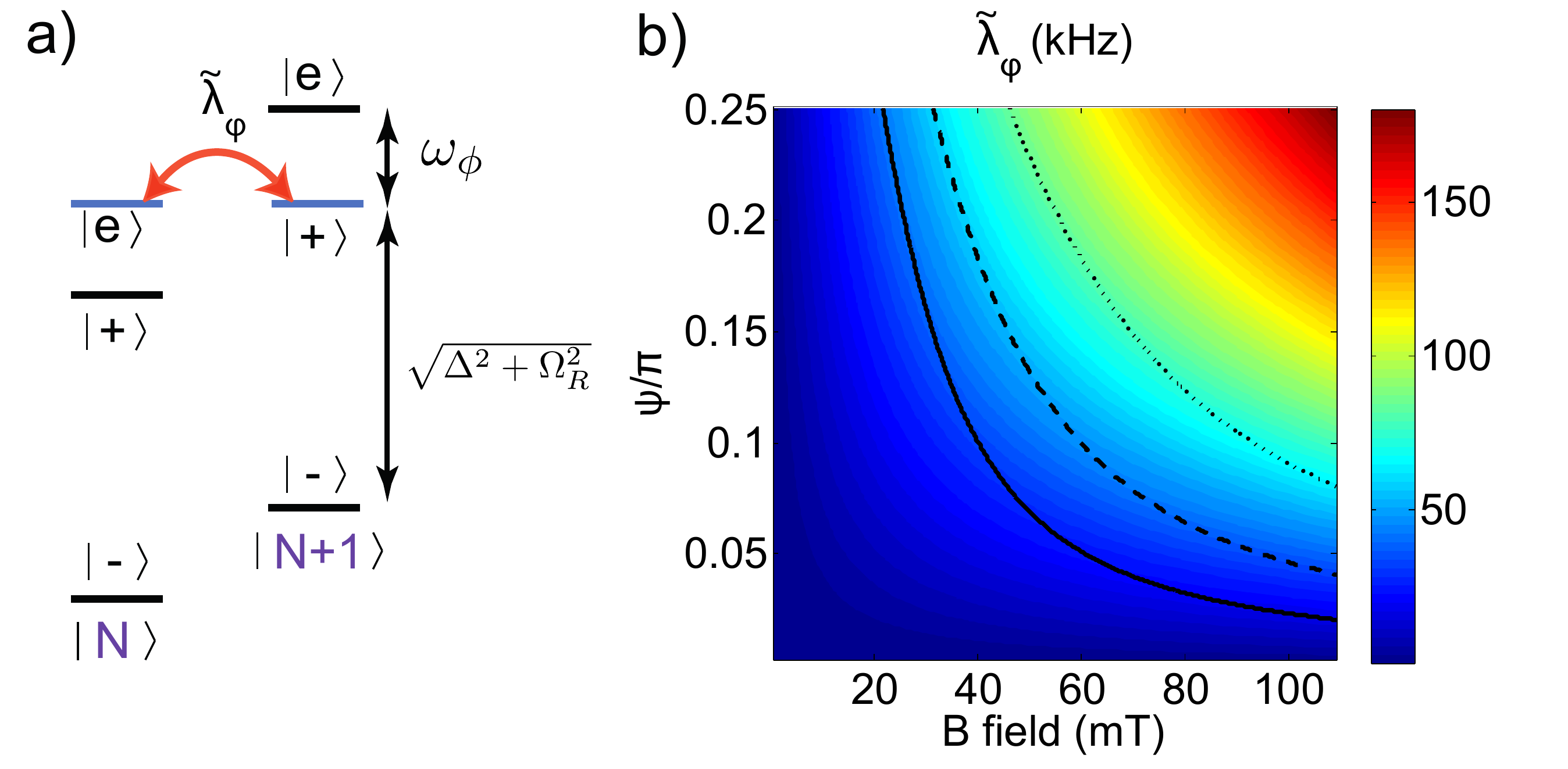}}
\caption{(a) Level diagram and coupling between the dressed spin states with rotational phonon numbers $N+1$ and $N$. 
(b) Coupling rate as a function of the $B$-field and $\Psi/\pi$ for  $\omega_\phi/(2\pi)$=5 MHz and a 20 nm diameter prolate particle. The black continuous, dashed and dotted lines are the $\psi$ parameter under resonant conditions as a function of the $B$-field for Rabi frequencies of 250, 500 and 1000 MHz respectively.}\label{coolingmap2}\label{coupling}
\end{figure}

\section{Quantization of the rotational mode} The rotational degree of freedom can be quantized like the center of mass mode.
For a small rotation of the diamond about the $y$ direction, the motional Hamiltonian can be linearized and takes the form
\begin{equation}\begin{aligned}
H_{\rm meca}=\frac{1}{2}I_y\omega_\phi^2\hat{\phi}^2+\frac{\hat{L}^2}{2I_y},
\end{aligned} \end{equation}
where $\omega_\phi$ is the rotational frequency, $I_y$ the moment of inertia with respect to the y axis and $\hat L$ the angular momentum.
It can now be written in the form of a harmonic oscillator with the two variables $\hat{L}$ and $\hat{\phi}$ .
In analogy to the canonical conjugate observable $\hat{X}$ and $\hat{P}$ of the center of mass mode, one can define
annihilation and creation operators $\hat{a}$ and $\hat{a}^\dagger$ such that
$\hat{\phi}=\phi_0(\hat{a}^\dagger +\hat{a})$, where $\phi_0=\sqrt{\hbar/(2I_y\omega_\phi)}$,
and $\hat{L}=I_y \hat{\dot\phi}=iL_0(\hat{a}^\dagger -\hat{a})$, where $L_0=\sqrt{\hbar I_y \omega_\phi/2}$
\footnote{Because of its multivalued character, quantization of the canonical conjugate angle and orbital angular momenta has been a controversial debate \citep{Kastrup}. Here, in the Paul trap, the angle is stabilized so that its value remains close to zero at all times, thus ensuring that the problem is analogous to the position and momentum degrees of freedom.}.
Fig. 1-b) shows the harmonic potential for a small angle $\phi$. 
A single quantum of motion will here have an angular extension $\phi_0$ inversely proportional to the square root of the moment of inertia $I_y$.

\section{Hamiltonian for rotational optomechanics} Having quantized the rotational mode, we now turn to the estimation of the coupling strength between the NV center spin and the rotational mode.
We take a NV center aligned to the $Z$ axis, which rotates with the diamond around the $y$ axis as depicted in figure \ref{setup}-a).
In the presence of a homogeneous transverse magnetic field along $x$, if the nanodiamond rotates, the projection of the spin component along the magnetic field is changed, thus providing a means to read-out the angular motion. A torque can then also be applied to the diamond via the NV spin.
The magnetic field dependent part of the Hamiltonian $\hat{H}_B= \gamma \vec{B} \cdot \hat{\vec{S}}$ describes the coupling of the spin $\hat{\vec{S}}$ to the transverse magnetic field $B$. Here $\gamma$ is the gyromagnetic ratio of the NV electron spin.
We take $\hat{S}_x$ and $\hat{S}_z$ the dimensionless spin operators along the $X$ and $Z$ rotating axes, we get 
$
\hat{H}_B= h \gamma B \left( \sin{\hat{\phi}} \hat{S}_z+\cos{\hat{\phi}} \hat{S}_x \right).
$
Considering only first order terms in $\hat{\phi}$, the magnetic Hamiltonian becomes
\begin{eqnarray}\hat{H}_B= h \gamma B \phi_0 \left( \hat{a}+\hat{a}^\dagger \right) \hat{S}_z + h \gamma B \hat{S}_x.
\end{eqnarray}
The total Hamiltonian of the system then reads :
\begin{equation}
\hat{H}=\hbar \omega_\phi \hat{a}^\dagger\hat{a}+ \hat{H}_{\rm NV}+h \lambda_\phi \left( \hat{a}+\hat{a}^\dagger \right) \hat{S}_z,\end{equation}
where $\hat{H}_{\rm NV}$ is the Hamiltonian of the NV spin without the opto-mechanical coupling term and where the single quantum of motional shift is given by
$
\lambda_\phi=\gamma B \phi_0.
$
The NV hamiltonian reads 
\begin{equation}\hat{H}_{\rm NV}=h D \hat{S}_z^2+h \gamma B \hat{S}_x + \hbar \Omega_R \hat{S}_y \cos(\omega t).\end{equation}
The first term $D S_z^2$ arises from the spin-spin coupling between the two electrons in the ground states and lifts the degeneracy between the $|{\pm 0}\rangle$ and the $|{\pm 1}\rangle$ spin states. For the NV center, $D=2.87$~GHz at room temperature. The second term, resulting from the transverse magnetic field, mixes the ground and excited electronic states $|{0}\rangle$, $|{\pm 1}\rangle$ into the mixed state $|{g}\rangle$, $|{d}\rangle$ and $|{e}\rangle$ that are presented bellow. The third term describes the coupling between the NV electronic spin and a microwave that is linearly polarised along the $y$ axis, at a frequency $\omega$ and Rabi frequency $\Omega_R$.
We will now diagonalize this hamiltonian and show how this configuration allow us to obtain the coupling between the spin and the rotational mode. Note that the choice for the magnetic field direction and microwave signal polarization is not critical and angles may be chosen to optimize the coupling rate \cite{ma2016quantum}. 

%\section{Diagonalisation and coupling strength}

In the absence of the microwave ($\Omega_R=0$) the eigenstates of $\hat{H}_{\rm NV}$ are the mixed state
$|{d}\rangle=\left(|{-1}\rangle-|{1}\rangle\right)/\sqrt{2}, 
|{g}\rangle=\cos \theta |{0}\rangle-\sin \theta |{b}\rangle, 
|{e}\rangle=\sin \theta |{0}\rangle+\cos \theta |{b}\rangle$
where $|{b}\rangle=\left(|{-1}\rangle+|{1}\rangle\right)/\sqrt{2}$, $\tan 2\theta=2 \gamma B/D$. The energies of these mixed states are $\omega_{e/g}=2\pi D\left(1\pm\sqrt{1+(2 \gamma B/D)^2}\right)/2$, $\omega_d=2\pi D$ and depends on the B field as shown fig. 1-c). \\

In the basis of these vectors, we have : $$ \hat{S_y} =\left( \cos \theta |{d}\rangle\langle{g}|+\sin \theta |{e}\rangle\langle{d}|-h.c.\right)/i.$$

In our case, we consider $\omega\sim\omega_{dg}=\omega_d-\omega_g\neq\omega_{ed}$ so that the microwave only drives the transition between the $|{g}\rangle$ and $|{d}\rangle$ mixed states, as is depicted in Fig. 1-a). 
We now move in a frame at the microwave frequency. In this frame, $\hat{H}_{\rm NV}$ reads
\begin{eqnarray}\hat{H}_{\rm NV}= \hbar/2 \begin{pmatrix}
-\Delta & 0 &0 \\
0 & \Delta &0 \\
0&0&\omega_{e'}\\
\end{pmatrix}+\hbar/2i \begin{pmatrix}
0 & -\Omega_R &0 \\
\Omega_R & 0 &0\\
0&0&0
\end{pmatrix},\end{eqnarray}
where $\Delta=\omega-\omega_{dg}$, $\omega_{e'}=\omega_e-(\omega+\omega_g+\omega_d)/2$ and the energy origin has been set to $(\omega_g+\omega_d)/2$.\\
The new eigenstates of this hamiltonian are now $|{e}\rangle$, 
$
|{+}\rangle= i \sin \psi |{g}\rangle+\cos \psi |{d}\rangle
$
and $|{-}\rangle= -i \cos \psi |{g}\rangle+\sin \psi |{d}\rangle, 
$
where $\tan 2 \psi=\Omega_R/\Delta$ and with $\omega_{+/-}=\pm\sqrt{\Delta^2+\Omega_R^2}/2$.
%\captionsetup[subfigure]{position=top, labelfont=bf,textfont=normalfont,singlelinecheck=off,justification=raggedright}
In the new eigenstate basis $|{+}\rangle$, $|{-}\rangle$, $|{e}\rangle$, the Hamiltonian can be approximated by a Rabi Hamiltonian :
\begin{eqnarray}\nonumber
\hat{H}&=&\hbar \omega_\phi \hat{a}^\dagger\hat{a}+ \hbar\omega_+ |{+}\rangle\langle{+}|+\hbar\omega_- |{-}\rangle\langle{-}|+\hbar\omega_{e'} |{e}\rangle\langle{e}|\\ 
&+&h \tilde{\lambda}_\phi \left( \hat{a}+\hat{a}^\dagger \right)\left(|{e}\rangle\langle{+}|+\rm{h.c.}\right),\label{eqHJC}
\end{eqnarray}
where $\tilde{\lambda}_\phi=\lambda_\phi \cos{\theta} \sin{\psi}$, which in turn can be reduced to a Jaynes-Cummings Hamiltonian under the condition that $\tilde{\lambda}_\phi \lesssim 10 |\omega_e-\omega_+|/(2\pi)$ and if we neglect the off-resonant terms \cite{braak2016analytical}.
Here, we set the microwave such that only the states $|{+,N}\rangle$ and $|{e,M}\rangle$ (with $M-N=\pm1$, $M$ and $N$ being the phonon numbers) are resonant, {\it i.e.} $\omega_{e'}-\omega_{+}= \omega_\phi$. In equation (\ref{eqHJC}), the other terms $|{+}\rangle\langle{-}|$ and $|{e}\rangle\langle{-}|$ are neglected in the rotating wave approximation.

Fig. \ref{spec}-b) depicts the $|{+,N}\rangle$ and $|{e,M}\rangle$ states in the resonant condition $\omega_{e'}-\omega_{+}= \omega_\phi$ : in the strong coupling regime, this Hamiltonian allows us to obtain a coherent exchange between rotational phonons and spin states at a rate $\tilde{\lambda}_\phi=\lambda_\phi \cos{\theta} \sin{\psi}$.\\

\section{Coupling rate}

In figure \ref{coupling}-b) the coupling rate $\tilde{\lambda}_\phi$ has been plotted as a function of the magnetic field and $\psi$ parameter, which depends on the microwave settings $\Omega_R$ and $\Delta$. It can be optimized by taking a resonant microwave ($\psi=\pi/4$) and a strong magnetic field. There is a practical limitation to accessing the area of this map however : as the magnetic field is increased, the resonance condition ($\omega_{e'}-\omega_+ = \omega_\phi$) requires a large splitting $\sqrt{\Delta^2+\Omega_R^2}$ between the $|{+}\rangle$ and $|{-}\rangle$ states. Since it is technically challenging to increase $\Omega_R$ above the GHz range \cite{fuchs2009gigahertz}, one will have to increase $\Delta$ therefore limiting oneself to lower $\psi$ values. The $\psi$ parameter needed to obtain the resonant condition as a function of the magnetic field given certain Rabi frequencies has been plotted above $\tilde{\lambda}_\phi$ in figure \ref{coupling}-b). These curves show that for all three Rabi frequencies, the optimum of the coupling rate is obtained with a resonant microwave. We now turn to the optimisation of $\tilde{\lambda}_\phi$ through the geometry of the particle.

\subsection{The role of the geometry}

Let alone the NV spin, using the rotational mode also has a number of advantages that was already emphasized by several groups working on optical tweezers \cite{kuhn2016full, Stickler, tongcangPRL}.
As we will show here, reducing the size of the particle to obtain a better charge to mass ratio and engineer its shape dramatically increases the rotational confinement.
The crucial parameters to attain the strong coupling regime can be extracted from the formulas $\phi_0=\sqrt{\hbar/2I_y\omega_\phi}$ and $\omega_\phi \sim \frac{Q S_\mu}{I_\mu}\frac{V_{ac}}{z_0^2}\frac{1}{\Omega} $. The angular frequency $\omega_\phi$ must be higher than the width of the electron spin resonance to achieve and coherent manipulation. Also, $\phi_0$ must be high enough in order for the phonon-photon coupling  rate $\tilde{\lambda}_\phi= \lambda_\phi\cos{\theta}\cos{\psi}$ to be higher than all decoherence rates.
Both $\omega_\phi$ and $\phi_0$ depend on parameters that are both intrinsic and extrinsic to the diamond. Because of the $\cos{\psi}$ term in $\tilde{\lambda}_\phi$, the rabi frequency $\Omega_R$ has to be as large as possible with no detuning $\Delta$. Here we take $\Omega_R=500$ MHz as a technical upper bond. This in turn limits the intensity of the B-field we can use while achieving the spin-phonon resonance to $\sim$30 mT. The other extrinsic parameters such as the Paul trap parameters $V_{\rm ac}$, $z_0$ and $\Omega$ can be increased to tune the frequency $\omega_\phi$. Technical limitations will however set an upper bound : the Paul trap should not be smaller than a few tens of microns and reaching a voltage higher than a few thousands volts at high frequencies will be challenging. The intrinsic parameters of the diamond particle are then the only parameters that can be tuned.

The frequency that must be attained is determined by both the quality of the diamond and the distance of the NV center from the surface, hence by the diameter $d$ of the considered diamond. However, one can see that decreasing the size of the diamond can considerably increase the trapping frequency. This is manifest in that $\frac{Q S_\mu}{I_\mu}=\frac{\sigma S S_\mu}{I_\mu}$ scales as $\frac{1}{d}$, provided the charge density on the diamond surface is independent on $d$.
Then, the factor $\frac{Q S_\mu}{I_\mu}$ depends strongly on the geometry. To obtain a high $\phi_0$, one also requires to have a low inertia momentum. This also points towards using small particles and highlights the relevance of the geometry.

In short, micron-size diamonds do not lend themselves easily to coherent manipulation since they are heavier, but they will retain the photophysical properties of bulk diamonds. Conversely, nanodiamonds have shorter coherence time than in the bulk but faster coupling rates can be reached. By tuning the aspect ratio of particles one can however find a compromise. 

\begin{figure}[ht!]
\centerline{\scalebox{0.3}{\includegraphics{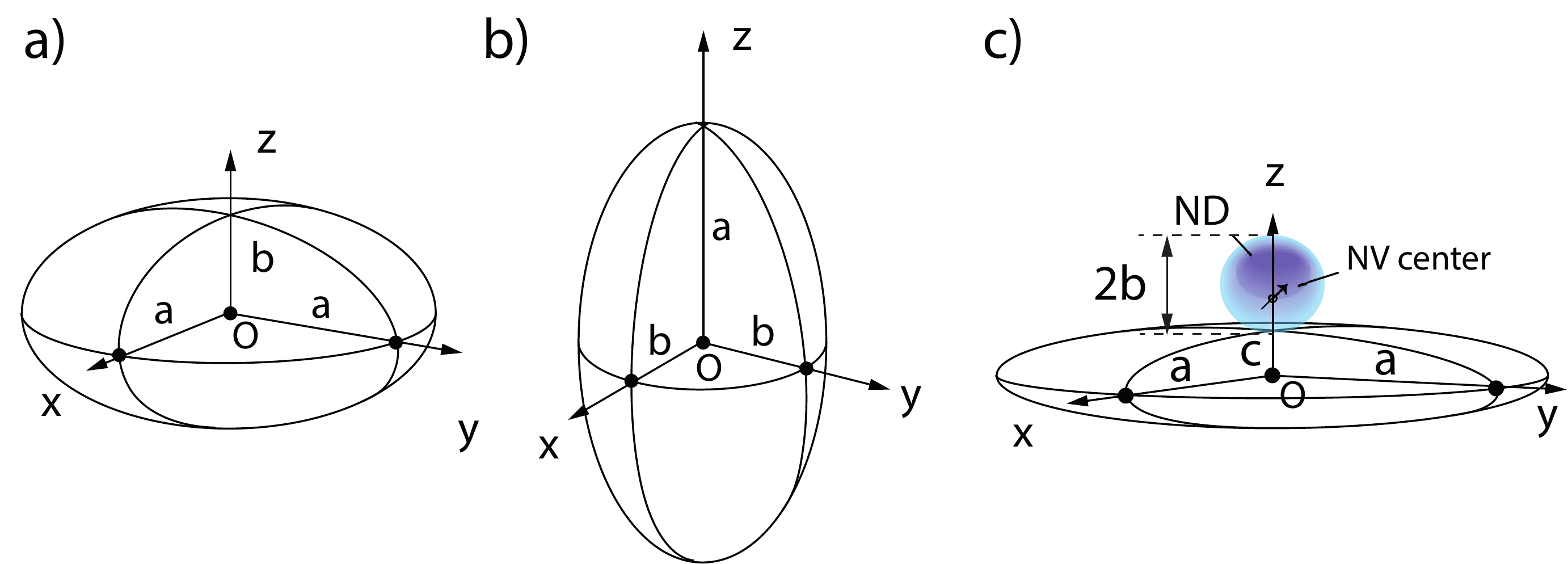}}}
\caption{Shapes of the proposed asymetric particles : a) oblate ellipsoid, b) prolate ellipsoid and c) composite particle formed by a diamond sphere deposited on a thin disk, approximated by an oblate ellipsoid.}\label{shapes}
\end{figure}

Three geometries have been envisioned and are depicted in figure \ref{shapes} : oblate and prolate ellipsoids, and composite particles formed by a diamond sphere deposited on a thin disk. 
This last shape enables choosing any material for the disk and thus optimize the charge to mass ratio and trapping frequency independent on the diamond. 
For all shapes, the $b$ and $a$ parameters always correspond to the minimum, resp. maximum, particle radii. 
In table \ref{geometries}, the trapping frequencies and moment of inertia are calculated for these shapes.
They are normalized with respect to the trapping frequencies ($\omega_0$) and moment of inertia ($I_0$) of a sphere with the same radius $b$. It is indeed important to compare particles with the same minimum radius to ensure that the NV properties (which depend crucially on their distance from the surface) are the same. The rotational frequency is also compared to the center of mass mode $\omega_{\rm com}$, for the same $b$ and for an aspect ratio $a/b=2.5$.  The trapping frequencies were calculated by integrating the torque over the surface of the ellipsoids, considering a homogeneous surface charge. Looking at table~\ref{geometries}, one sees that particles with a higher asymmetry and spatial extent experience a higher rotational trapping frequency. They however also have a greater moment of inertia, which will reduce the coupling $\lambda_\phi$. To both increase the trapping frequency and reduce the inertia momentum without reducing the size of the particle, the proposed composite particle comprising a spherical diamond of size $b$ within or deposited on a thinner disk of silica allows to considerably increase the confinement of the particle with a moment of inertia much smaller than with simpler shapes, as can be seen in table \ref{geometries}.

\begin{table}[ht!]
\centerline{\scalebox{1}{\begin{tabular}{|l|c||c|c|c|c|}
\hline
particle type & $c/b$ & $\omega_{com \rm}/\omega_0$  & $\omega_{\phi}/\omega_{0}$ & $\omega_{\phi}/\omega_{com \rm}$ & $I_y/I_0$  \\
\hline
sphere				& - & 1 & 0 & 0 & 1\\
\hline
oblate ellipsoid  	& - & 0.64 & 1.8 & 2.9 & 23\\
\hline
prolate ellipsoid 	& - & 0.83 & 2.3 & 2.8 & 9\\
\hline
composite			& 0.125 & 2.8 & 19 & 5.3 & 2.4\\
\hline
composite			& 0.0625 & 3.3 & 27.6 & 6.3 & 1.2\\
\hline
\end{tabular}}}
\caption{
Comparison of the mechanical parameters of different particles shapes, as shown in fig. \ref{shapes} for the same $b$, for the same aspect ratio $a/b=2.5$ and identical surface charge density. $\omega_0$ and $I_0$ are the secular frequency of the center of mass and the moment of inertia of a sphere with radius $b$ respectively. For each considered particles, $\omega_{com \rm}$ and $\omega_{\phi}$ are the secular frequencies of the center of mass and the rotational modes respectively, $I_y$ is the moment of inertia.}\label{geometries}
\end{table}

We note that, even though advances have been made in engineering the shape of nanodiamonds, the proposed ellipsoidal particles are an approximation of the particle shapes of  \cite{appel2016fabrication,andrich2014engineered,shallow25nm}.
Reactive ion etching (RIE) \cite{pilliers111, shallow25nm} is however quite adapted for rotational optomechanics, since diamonds can be engineered to form nano-pillars that are close to prolate ellipsoidal particles.

\captionsetup[subfigure]{position=top, labelfont=bf,textfont=normalfont,singlelinecheck=off,justification=raggedright}
\begin{figure}[ht!]
\centering
\subfloat[\label{coolingB20nm}]{\includegraphics[width=0.24\textwidth]{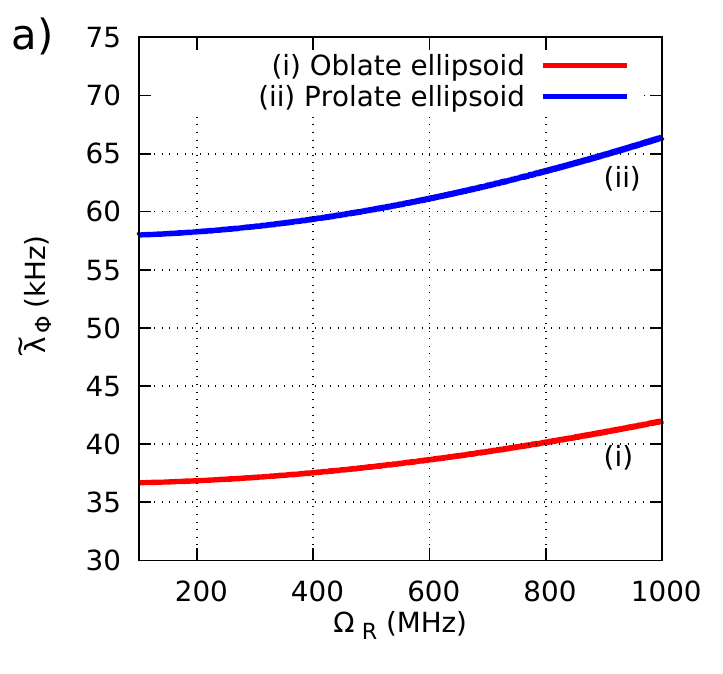}}
\subfloat[\label{coolingB80nm}]{\includegraphics[width=0.24\textwidth]{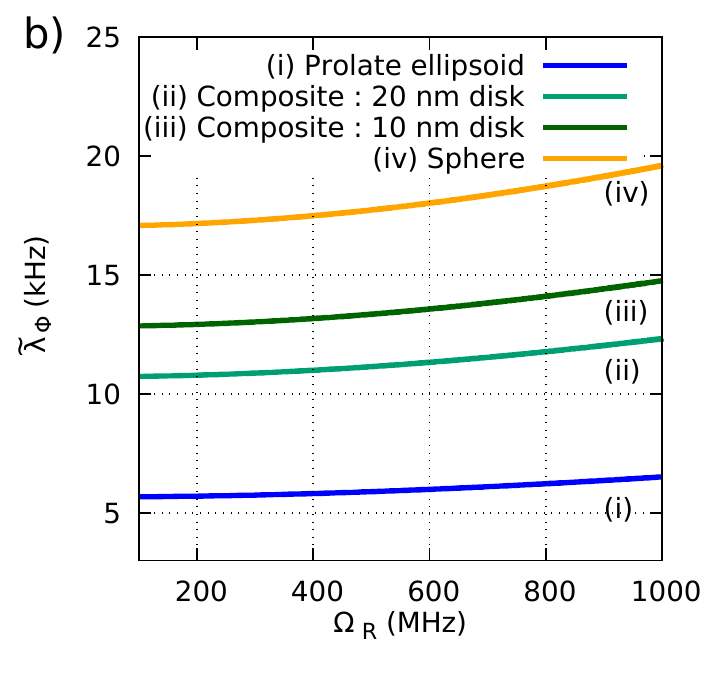}}
\caption{a) Coupling rate $\tilde{\lambda}_\phi$ for nanodiamonds of different shapes with a) a radius $b=20$ nm for the prolate and oblate ellipsoids with a rotational confinement $\omega_\phi=5$ MHz and b) with $b=80$ nm and a rotational confinement $\omega_\phi=0.5$ MHz. The microwave is set at resonnance with a varying Rabi frequency and the magnetic field is tuned to obtain resonant conditions. For example at $\Omega_R=$500 MHz we have B$\sim$30 mT. The aspect ratio of the proposed particles is $a/b=2.5$ for all particles and $c/b=1/8$, $1/16$ for the composite 20~nm and 10~nm disks respectively. The coupling rate with a zero-mass disk ({\it i.e.} for a sphere) is plotted as a limit for such particles (trace iv).}\label{coolingB} 
\end{figure}

Let us now estimate the spin-phonon coupling rate for the above particule geometries.
We compare two different particle sizes. 
In figure \ref{coolingB}, the coupling rate $\tilde{\lambda}_\phi$ is plotted as a function of the Rabi frequency for the particle geometries described above, such as oblate, prolate or composite particles.
%In figure \ref{coolingB}-b), two cases for particles of different sizes are studied, depending on the minimum bulk radius around the center of the particle as this should determine the properties of the most buried NV centers.
Fig. \ref{coolingB}-a) shows the coupling rate for particles with a radius $b=20$ nm and and aspect ratio of $a/b=2.5$ as a function of the Rabi frequency and with the magnetic field tuned to achieve resonant conditions.
The coupling increases with the Rabi frequency as expected since the higher it is, the higher can the magnetic field be while still fulfilling the resonance condition. Here for a Rabi frequency $\Omega_R=500$ MHz and a magnetic field $B\sim 30$ mT $\tilde{\lambda}_\phi$ ranges between 35 to 60 kHz. In fig. \ref{coolingB}-b) the considered particles are chosen to have a radius $b=80$ nm. Due to the high mass, the coupling rate for a prolate ellipsoid is then smaller, barely exceeding 5 kHz. It can however be increased using a composite particle made out of a silica pancake-like shape with a nanodiamond deposited on top. The coupling strength will then depend crucially on how thin can the disk of the composite particle be. 

\subsection{The total number of charges}

In order to reach significant trapping frequencies, the surface charges on the nanodiamond must be large enough, which in turn may yield charge fluctuations that will affect the NV spin coherence time.

First, in order to obtain an order of magnitude for the needed total surface charge, we compute the trapping frequency for the needle trap. The curvature of and distance between the needles determine the confinement and the potential depth in both the radial and axial planes \cite{Deslauriers}.  The axial angular frequency $\omega_z$ of the harmonic pseudo-potential is given by 
$
\omega_z=|Q_{\rm tot}| V_{\rm ac} \eta/\sqrt{2} m \Omega z_0^2,
$
where $V_{\rm ac}$ is the peak to peak voltage applied between the electrodes and the far distance surrounding ground, 
$m$ is the mass of the trapped particle and $\eta$ the efficiency factor
that accounts for the reduction in the trap potential as compared 
to an analogous quadrupole trap with hyperbolic electrodes.
$\Omega/2\pi$ is the trapping frequency and $Q_{\rm tot}$ is the total excess charge of the particle. 
One can then deduce the total surface charge needed to reach the two considered rotational frequencies given for the Fig. 3 a) and b) in the main text.
Taking a prolate nanodiamond particle with an aspect ratio of 2.5, we find, looking at table 1, the rotational mode to be about 3 times larger than the center of mass radial mode.
We thus get  $$|Q_{\rm tot}|= \frac{2\sqrt{2} m \Omega z_0^2}{3 \omega_\phi \eta V_{\rm ac}},$$
We then take a distance between the electrodes of 10 $\mu$m, an efficiency parameter $\eta=0.3$, a voltage $V_{ac}$=5000 V and a trapping frequency of 5 MHz. 
In order to reach a trapping frequency of 0.5 MHz and with $b=80$ nm, we find that at least 60 elementary charges must be on the diamond surface. 
These charges can originate from remaining $sp^2$ layers or can be generated {\it in situ} using UV light.

\section{Decoherence sources\label{limitations}}

The so-called strong coupling regime is reached if the spin-phonon coupling rate exceeds the decoherence rates of both the spin and the considered mechanical oscillator mode. For the spin, this strong coupling condition translates to $T_1,T_2 \gg 1/\tilde{\lambda}_\phi $, where $T_2^{-1}=(2T_1)^{-1}+(T_2^*)^{-1}$, and $(T_2^*)^{-1}$ is the inhomogeneous decoherence rate due to the coupling between the NV spin to a nuclear spin bath or due to charge fluctuations.

\subsubsection{Nuclear spin bath}
For very shallow (5 nm deep) NV centers, $T_1$ generally ranges from several hundreds of microseconds to milliseconds \cite{shallow5nm}, the main constraint for reaching the strong coupling regime is therefore the $T_2^*$ time. Studies using dynamical decoupling sequences show that NV centers within 50 nm diameter nanodiamonds synthetized through RIE \cite{shallow25nm} of CVD-grown diamonds should reach $T_2$ times of up to 200 $\mu s\rm$ using $^{12}$C isotopically engineered diamond \cite{balasubramanian2009ultralong,maze2008electron}. It was indeed shown that the main source of degradation of the $T_2$ is the $^{13}$C nuclear spin bath. For particles with a 20 nm minimum bulk radius, a $T_2\sim 150$ $\mu$s would already allow entering the strong coupling regime. For particles with a 80nm radius, $T_2$ is expected to be closer to bulk values, and can attain $T_2\sim 1.8$ ms \citep{balasubramanian2009ultralong}. Looking at Fig. 3-b), here $T_2\gtrsim 1$ ms would already be sufficient to reach the strong coupling regime.

\subsubsection{Charge fluctuations}
Aside from assisting the rotational coupling, the charges may have detrimental effects on the electronic spin.
First, surface charges have been shown to affect the NV$^-$ to NV$^0$ conversion. For the considered diamond diameters ($>$ 40 nm), this effect is however not significant \cite{Rondin2}. 
It was shown that the dominant source of electric noise under zero-magnetic field is related to the optical illumination of the NV center for centers deeper than 100 nm, and remains significantly smaller than the magnetic noise for the magnetic field tens of mT used in the proposed rotational opto-mechanics \cite{Jamonneau}.
The electric field noise contribution is expected to increase significantly only when the distance from the surface is below 10 nm, owing to the close vicinity of fluctuating charges lying on the diamond surface \cite{kim}. 
If in the end surface charge fluctuations still cause decoherence to the NV spin, one could also expect that they will be modified, and possibly reduced, by the Paul trap. Due to the motional instability of the small electron masses, the trap may apply an effective outward force, potentially stabilizing the surface electrons and thus reducing the electric noise. This last conjecture however remains to be checked experimentally or theoretically with a model dealing with the actual surface termination.

\subsubsection{Spin polarisation}
In this paper, we did not touch upon phonon and spin state preparation and read-out, which can be done using the present coupling scheme. 
One foreseeable concern is the transverse B-field of NV centers, which degrades the spin polarisation and read-out efficacy 
by inducing spin mixing of the ground and excited spin states \cite{tetienne2012magnetic}.
An optimum transverse magnetic field should thus be found for initialization and read-out of the spin state.

The strong coupling condition itself is however unaffected by such spin mixing as it does not rely on the spin-selective non-radiative de-excitation.
For the full protocol including rotational mode and spin preparation, spin-phonon coupling and final phonon state read-out, a sequence with an optimized time-dependent transverse magnetic field seems necessary and will be dealt with in a forthcoming paper.

\section{Conclusion}

To conclude, we have shown how to benefit from the original spin properties of the NV center to obtain a rotational opto-mechanical coupling using a nanodiamond levitating in a Paul trap. We show how to enter the strong coupling regime for different particle sizes and shapes.
This quantum opto-mechanical approach is promising in that it uses a scattering-free trapping of nanodiamonds at room temperature and circumvents the necessity to employ very strong magnetic gradients \cite{Rabl, yin}. In the longer run, this platform will enable efficient quantum control of macroscopic oscillators, paving the way towards Schr${\rm \ddot o}$dinger cat states where the NV spin is entangled with the collective rotational motion of millions of atoms \cite{yin,yin2013optomechanics}. This novel architecture will furthermore open opportunities for studying fundamental phenomena in quantum optics and establish building blocks of future quantum-based technologies. \\

%Compared to other proposals that use strong magnetic field gradients for motional read-out and intense light fields for the trapping \cite{yin2013optomechanics}, the proposed scheme enables entering the strong coupling regime for macroscopic oscillators with only modest magnetic and light field amplitudes. \\

\noindent \textit{Note : During the preparation of this manuscript, complementary work by Y. Ma and coworkers was
presented in \citep{ma2016quantum}.}

\section*{Acknowledgements}
We would like to acknowledge fruitful discussions with Peter Rabl. 
This research has been partially funded by the French National Research Agency (ANR) through the project SMEQUI.

\end{document}